\documentclass[twocolumn,showpacs,preprintnumbers,amsmath,amssymb]{revtex4}
\usepackage{bm}
\begin{document}
\preprint{}
\title{A minimal three-flavor model for neutrino
oscillation based on superluminal property}
\author{Guang-Jiong Ni}
\affiliation{Department of Physics, Fudan University, Shanghai, 200433, China\\
Department of Physics, Portland State University, Portland, OR 97207 USA}
    \altaffiliation[ ]{E-mail address: gj\_ni@Yahoo.com}
\date{13 May 2003}
\begin{abstract}
Recent experimental research progress of solar neutrino and neutrino oscillation reveals
that neutrinos do have nonzero mass, which poses a serious challenge to theoretical
physics. Two symmetries, the common invariance under the space-time inversion and the
invariance or violation to maximum under the space inversion are analyzed for Dirac
equation or the equation for superluminal neutrino respectively. Then a simple
three-flavor model containing only one parameter is proposed and the relations among
neutrino oscillation,parity violation and superluminal property are emphasized.
\end{abstract}
 \maketitle

Since the experimental verification of neutrino oscillation, i.e., the mutual
transformation among three flavors of neutrinos ( $\nu_e$, $\nu_\mu$, and $\nu_\tau$ ) by
Super-Kamiokande Collaboration [1] in 1998 and further by SNO Collaboration in 2002 [2],
a common cognition that neutrinos have nonzero mass has been accepted by physics
community. This is because for a two-neutrino oscillation model, the probability for a
neutrino produced in flavor state $a$ to be observed in flavor state $b$ after travelling
a distance L through a vacuum is [1]:
\begin{equation}\label{eq:1}
  P(\nu_a\rightarrow\nu_b)=\sin^2 2\theta\sin^2\left(\frac{1.27\triangle m^2({\rm eV^2})L({\rm km})}{E_\nu
  ({\rm GeV})}\right)\,,
\end{equation}
where $E_\nu$ is the neutrino energy (in GeV), $\theta$ is the mixing angle between the
flavor eigenstates and the mass eigenstates, and $\triangle m^2$ is the mass square
difference of the neutrino eigenstates. Obviously, once $\triangle m^2=0$, there will be
no neutrino oscillation at all ( see the excellent books [3, 4]).

However, since the discovery of parity violation by Lee-Yang, Wu et al in 1956-57 and the
ingenious experiment by Goldhaber et al [5], it is concluded that neutrino is always
left-handed ($\nu_{_L}$) while antineutrino right-handed ($\overline{\nu}_{_R}$). Once
they have mass and so are Dirac particles moving in a speed $u$ less than light-speed
$c$, then an observer moving in a parallel velocity $v>u$ would see a neutrino becoming
right-handed ($\nu_{_R}$) and an antineutrino left-handed ($\overline{\nu}_{_L}$). But
neither $\nu_{_R}$ nor $\overline{\nu}_{_L}$ exists in experiments. Hence the evidence
for neutrino mass poses a serious challenge to theoretical physics [6].

There are some approaches to make neutrino massive [3, 4]. One approach assumes that the
$\nu_{_R}$ interacts much more weakly than other particles and so is unobserved. the
second way involves a theory of Majorana neutrino (in which neutrino and antineutrino can
be identical) in combination with so-called seesaw mechanism to allow the extremely heavy
$\nu_{_R}$ existing for a very short time interval and only let $\nu_{_L}$ to be
observed. I have serious doubts about the above approaches, based on the criterion of
beauty. As Dirac stressed: ``Physical laws should have mathematical beauty" (see [7]
foreword xiii and p.257)." ``It seems to be one of the fundamental features of nature
that fundamental physical laws are described in terms of great beauty and power. As time
goes on it becomes increasingly evident that the rules that the mathematician finds
interesting are the same that Nature has chosen" (quoted from [8]). It seems to me that
there is another very simple and interesting approach to make neutrino massive while in
the meantime respect to the parity violation, i.e., the permanently longitudinal
polarization of neutrinos. Based on a series of works [9-15], I will discuss the equation
and property of a subluminal or superluminal particle before a simple three-flavor model
for neutrino oscillation can be proposed.

Beginning from the famous Dirac equation, I wish to emphasize its essential beauty being
hidden in two symmetries. Let us write it in a form of two coupled equations of
two-component spinors $\varphi_{_D}(\bm x,t)$ and $\chi_{_D}(\bm x,t)$, the hidden
particle field and antiparticle field[16]:
\begin{equation}\label{eq:2}
  \left\{\begin{array}{ll}
  i\,\hbar\frac{\displaystyle\partial}{\displaystyle\partial t}\;\varphi_{_D}=i\,c\,\hbar\,
  \bm\sigma\cdot\bm\nabla\chi_{_D}+m_{_0}\,c^2\,\varphi_{_D},\\[3mm]
  i\,\hbar\frac{\displaystyle\partial}{\displaystyle\partial t}\;\chi_{_D}=i\,c\,\hbar\,
  \bm\sigma\cdot\bm\nabla\varphi_{_D}-m_{_0}\,c^2\,\chi_{_D},
  \end{array}\right.
\end{equation}
Where $\bm\sigma$ are $2\times 2$ Pauli matrices. The reason why we prefer this form
rather than the covariant form of four-component spinor
$\psi_{_D}=\left(\begin{array}{cc}\varphi_{_D}\\\chi_{_D}\end{array}\right)$ lies in the
fact that the symmetries will become explicit as follows.

First, Eq. (2) is invariant under the (newly defined) space-time inversion ($\bm
x\rightarrow-\bm x,t\rightarrow-t$)with transformation (see [16]):
\begin{equation}\label{eq:3}
  \varphi_{_D}(-\bm x, -t)\rightarrow\chi_{_D}(\bm x,t),\quad \chi_{_D}(-\bm x, -t)
  \rightarrow\varphi_{_D}(\bm x,t).
\end{equation}
When $|\varphi_{_D}|>|\chi_{_D}|$, it is a particle; when
$|\chi^c_{_D}|>|\varphi^c_{_D}|$, it shows up as an antiparticle. Next, after we define
the hidden left-handed and right-handed fields as:
\begin{equation}\label{eq:4}
  \xi_{_D}=\frac{1}{\sqrt{2}}(\varphi_{_D}+\chi_{_D}),\quad\eta_{_D}=\frac{1}{\sqrt{2}}(\varphi_{_D}-\chi_{_D}).
\end{equation}
Eq.~(2) is recast into:
\begin{equation}\label{eq:5}
  \left\{\begin{array}{ll}
  i\,\hbar\frac{\displaystyle\partial}{\displaystyle\partial t}\;\xi_{_D}=i\,c\,\hbar\,
  \bm\sigma\cdot\bm\nabla\xi_{_D}+m_{_0}\,c^2\,\eta_{_D},\\[3mm]
  i\,\hbar\frac{\displaystyle\partial}{\displaystyle\partial t}\;\eta_{_D}=-i\,c\,\hbar\,
  \bm\sigma\cdot\bm\nabla\eta_{_D}+m_{_0}\,c^2\,\xi_{_D}\,.
  \end{array}\right.
\end{equation}
When  $|\xi_{_D}|>|\eta_{_D}|$, it is a left-handed particle; when
$|\eta_{_D}|>|\xi_{_D}|$, it becomes right-handed explicitly. Now we see that Eq.~(5) is
invariant under the pure space-inversion ($\bm x\rightarrow-\bm x,t\rightarrow t$) with
transformation:
\begin{equation}\label{eq:6}
  \xi_{_D}(-\bm x, t)\rightarrow\eta_{_D}(\bm x,t),\quad \eta_{_D}(-\bm x, t)
  \rightarrow\xi_{_D}(\bm x,t).
\end{equation}

The solution of either (2) or (5) gives the wave-function(WF) describing a freely moving
Dirac particle:
\begin{equation}\label{eq:7}
  \varphi_{_D}\sim\chi_{_D}\sim\xi_{_D}\sim\eta_{_D}\sim exp\,[i\,(\bm p\cdot\bm
  x-Et)/h]\,,
\end{equation}
with the momentum $\bm p$ and energy $E$ satisfying the kinematical relation in special
relativity (SR):
\begin{equation}\label{eq:8}
  E^2=p^2 c^2+m^2_0 c^4.
\end{equation}

Usually the Dirac equation was regarded as an outcome of combination of quantum mechanics
(QM) with the theory of SR. However, actually, what we described above should be viewed
as a derivation of Eq.(8) in SR from QM by injecting two symmetries, i.e., in our
opinion, QM and SR are identical in essence[16]. Just as Gross stressed: ``the primary
lesson of physics in 20th century is that the secret of nature is symmetry".[8]

Based on Dirac's experience together with a belief that a particle is always not pure and
there is no exception to this rule, we manage to establish the equation for a
superluminal neutrino as:
\begin{equation}\label{eq:9}
  \left\{\begin{array}{ll}
  i\,\hbar\frac{\displaystyle\partial}{\displaystyle\partial t}\;\varphi=i\,c\,\hbar\,
  \bm\sigma\cdot\bm\nabla\chi+m_s\,c^2\,\chi,\\[3mm]
  i\,\hbar\frac{\displaystyle\partial}{\displaystyle\partial t}\;\chi=i\,c\,\hbar\,
  \bm\sigma\cdot\bm\nabla\varphi-m_s\,c^2\,\varphi,
  \end{array}\right.
\end{equation}
or equivalently:
\begin{equation}\label{eq:10}
  \left\{\begin{array}{ll}
  i\,\hbar\frac{\displaystyle\partial}{\displaystyle\partial t}\;\xi=i\,c\,\hbar\,
  \bm\sigma\cdot\bm\nabla\xi-m_s\,c^2\,\eta,\\[3mm]
  i\,\hbar\frac{\displaystyle\partial}{\displaystyle\partial t}\;\eta=-i\,c\,\hbar\,
  \bm\sigma\cdot\bm\nabla\eta+m_s\,c^2\,\xi,
  \end{array}\right.
\end{equation}
with Eq.~(4) without subscript $D$. We see that while Eq.~(9) remains invariant under the
space-time inversion (3), Eq. (10) is violated (to maximum) under the space inversion (6)
due to opposite signs in the mass term. A similar plane-wave solution like (7) now
yields:
\begin{equation}\label{eq:11}
  E^2=p^2 c^2-m^2_s c^4,
\end{equation}
which in turn implies a kinematical relation being:
\begin{equation}\label{eq:12}
  p=\frac{m_s u}{\sqrt{\frac{u^2}{c^2}-1}},\quad E=\frac{m_s
  c^2}{\sqrt{\frac{u^2}{c^2}-1}},
\end{equation}
where the particle velocity $u$ is defined as the group velocity $u_g$ of wave (
$E=\hbar\omega, p=\hbar k$):
\begin{equation}\label{eq:13}
  u=u_g=\frac{\displaystyle d\omega}{\displaystyle dk}
  =\frac{\displaystyle dE}{\displaystyle dp}=\frac{c^2 p}{E}>c.
\end{equation}
Hence, it is a superluminal particle (or tachyon) and $m_s$ is called the proper mass or
tachyon mass. On the other hand, the phase velocity $u_p$ of wave reads:
\begin{equation}\label{eq:14}
  u_p=\omega /k=E/p<c,
\end{equation}
because Eq.~(11) (like (8)) always imposes a constraint on the product of  $u_g$ with
$u_p$:
\begin{equation}\label{eq:15}
  u_g u_p=c^2.
\end{equation}
Some consequences of the above theory are as follows:

(a) As discussed in Refs[15, 16], a particle and its antiparticle obey the same equation,
e.g., Dirac Eq.~(2) or neutrino Eq.~(9). The WF for particle under the condition
$|\varphi|>|\chi|$ has positive-energy $E>0$ whereas that for antiparticle under the
condition $|\chi_c|>|\varphi_c|$ has negative-energy $E<0$ (actually it has a
positive-energy $E_c=-E>0$, see [15,16]).

(b) In contrast to Dirac Eq.~(5), the neutrino Eq.~(10) only allows solutions of
$\nu_{_L}$ (with $|\xi|>|\eta|$) and $\overline{\nu}_{_R}$ (with $|\eta_c|>|\xi_c|$),
other two kinds of solution --- $\nu_{_R}$ and $\overline{\nu}_{_L}$ --- are strictly
forbidden. This is a direct consequence of parity violation. And because now $\nu_{_L}$
and $\overline{\nu}_{_R}$ have velocity $u>c$, their polarizations (helicities) in one
frame can be maintained in another frame with relative velocity $v(<c)$ between two
observers.

(c) A comparison between Eq.~(8) and (11) reveals that if $m^2_0=-m^2_s$, we may write
$m_0=im_s$ (with real $m_s$) and it was often said that the kinematical relation for
superluminal particle could be viewed as some analytic continuation of that for
subluminal particle.

Experimentally, in the measurements like that of tritium beta decay, physicists did find
that in the kinematical relation of neutrino
\begin{equation}\label{eq:}
  E^2=p^2 c^2+m^2_\nu c^4,\nonumber
\end{equation}
the mass square seems negative with large uncertainty:
\begin{equation}\label{eq:16}
  \begin{array}{ll}m^2(\nu_e)=-2.5\pm 3.3~{\rm eV^2},\\[1mm]
  m^2(\nu_\mu)=-0.016\pm0.023~{\rm MeV^2},\end{array}
\end{equation}
as summarized in the particle table by 2000 [17]. (See Eqs.~(29) and (30).)

(d)While Dirac Eq.~(2) and superluminal Eq.~(9) share the common basic symmetry under the
space-time inversion (3), they have different symmetry under the space inversion (6):
Dirac Eq.~(5) remains invariant whereas superluminal Eq.~(10) shows maximum violation,
which should be viewed as some antisymmetry rather than asymmetry. In short, either Dirac
equation or superluminal equation is destined for specific realization of symmetries.

We are now in a position to discuss the neutrino oscillation, generalizing the Eq.~(10)
to a model for two flavors, $\nu_e$ and $\nu_\mu$. The coupled equations are assumed to
be ($m_e=m_s(\nu_e), m_\mu=m_s(\nu_\mu), \hbar=c=1$):
\begin{equation}\label{eq:17}
  \left\{\begin{array}{llll}
  i\,\dot{\xi}_e=i\bm \sigma\cdot\bm \nabla\xi_e-m_e\eta_e-\delta\,\eta_\mu,\\[2mm]
  i\,\dot{\eta}_e=-i\bm \sigma\cdot\bm \nabla\eta_e+m_e\xi_e+\delta\,\xi_\mu,\\[2mm]
  i\,\dot{\xi}_\mu=i\bm \sigma\cdot\bm \nabla\xi_\mu-m_\mu\eta_\mu-\delta\,\eta_e,\\[2mm]
  i\,\dot{\eta}_\mu=-i\bm \sigma\cdot\bm \nabla\eta_\mu+m_\mu\xi_\mu+\delta\,\xi_e\,.
  \end{array}\right.
\end{equation}
Notice that: First, the additional terms with coupling constant $\delta$ do not spoil the
symmetry in (10): Eq.~(17) still remains invariant under the space-time inversion (3) and
in the meantime violates the space-inversion symmetry (6) to maximum. Second, like
Eq.~(10), Eq.~(17) has the probability density being:
\begin{equation}\label{eq:18}
  \rho=\xi^\dag_e\xi_e+\xi^\dag_\mu\xi_\mu-\eta^\dag_e\eta_e-\eta^\dag_\mu\eta_\mu,
\end{equation}
which is not positive definite. $\int\rho d\bm x=1$ means the oscillation occurred
between left-handed $\nu_e$ and $\nu_\mu$ ($|\xi|>|\eta|$), while $\int\rho d\bm x=-1$
means that between right-handed $\overline{\nu}_e$ and $\overline{\nu}_\mu$
($|\eta_c|>|\xi_c|$).

Third, using the trick and notation like that by Sassaroli[18], we find two
eigen-energies of Eq.~(17) as:
\begin{equation}\label{eq:19}
  \begin{array}{lll}
  E^2_1=p^2-m^2_1,\quad E^2_2=p^2-m^2_2,\\[2mm]
  m_{1,2}=\frac{1}{2}[(m_e+m_\mu)\pm R],\\[2mm]R=\sqrt{(m_\mu-m_e)^2+4\delta^2}.
  \end{array}
\end{equation}
Denoting $\Lambda=\frac{\displaystyle m_\mu-m_e+R}{\displaystyle 2\delta}$,
$\sin\theta=1/\sqrt{1+\Lambda^2}$, $\cos\theta=\Lambda/\sqrt{1+\Lambda^2}$,
$\nu_\mu=\left(\begin{array}{ll}\xi_\mu\\\eta_\mu\end{array}\right)$,
$\nu_e=\left(\begin{array}{ll}\xi_e\\\eta_e\end{array}\right)$, we can express the mass
eigenstates as (see Appendix):
\begin{equation}\label{eq:20}
  \left(\begin{array}{cc}\nu_1\\\nu_2\end{array}\right)
  =\left(\begin{array}{cc}\cos\theta &\sin\theta\\-\sin\theta
  &\cos\theta\end{array}\right)
  \left(\begin{array}{cc}\nu_\mu\\\nu_e\end{array}\right)\,.
\end{equation}
Hence the time evolution of flavor amplitude is given by:
\begin{equation}\label{eq:21}
  \left(\begin{array}{cc}C_\mu(t)\\C_e(t)\end{array}\right)
  =\left(\begin{array}{cc}\cos\theta &-\sin\theta\\\sin\theta
  &\cos\theta\end{array}\right)
  \left(\begin{array}{cc}C_1(0)e^{-iE_1 t}\\C_2(0)e^{-iE_2 t}\end{array}\right)\,.
\end{equation}
Suppose that at $t=0$, an electron-neutrino is produced:
\begin{equation}\label{eq:22}
  C_\mu(0)=0,\quad C_e(0)=1\,.
\end{equation}
We obtain the oscillation probability of $\nu_e$ to $\nu_\mu$ after a time $t$ being:
\begin{equation}\label{eq:23}
  P(\nu_e\rightarrow\nu_\mu)=|C_\mu(t)|^2=\sin^2 2\theta\sin^2\left(\frac{(m^2_1-m^2_2)t}{4E}\right),
\end{equation}
while
\begin{equation}\label{eq:24}
  |C_e(t)|^2=1-\sin^2 2\theta\sin^2\left(\frac{(m^2_1-m^2_2)t}{4E}\right),
\end{equation}
where an ultrarelativistic approximation $p\gg m, p\sim E$ has been used. Eq.~(23) is
exactly Eq.~(1) ($t\sim L$) except that now $\triangle m^2=m^2_1-m^2_2$ is referred to
the square difference of tachyon mass for superluminal neutrino rather than that of rest
mass for sublumonal neutrino.

Since the experimental data show that the oscillation amplitude tends nearly to maximum,
we would expect in the above two-flavor model that $\theta\rightarrow \pi/4$, or
$\Lambda\rightarrow 1$, i.e., $\delta\gg(m_\mu-m_e)$. But a vanishing of $m_{\mu}$ and
$m_e$ is not allowed because it would eliminate the oscillation entirely.

Interesting enough, we can keep only one coupling parameter in a three-flavor model by
first generalizing (17) without mass term into:
\begin{equation}\label{eq:25}
  \left\{\begin{array}{llllll}
  i\,\dot{\xi}_e=i\bm \sigma\cdot\bm \nabla\xi_e-\delta\eta_\mu-\varepsilon\,\eta_\tau,\\[2mm]
  i\,\dot{\eta}_e=-i\bm \sigma\cdot\bm \nabla\eta_e+\delta\xi_\mu+\varepsilon\,\xi_\tau,\\[2mm]
  i\,\dot{\xi}_\mu=i\bm \sigma\cdot\bm \nabla\xi_\mu-\lambda\eta_\tau-\delta\,\eta_e,\\[2mm]
  i\,\dot{\eta}_\mu=-i\bm \sigma\cdot\bm \nabla\eta_\mu+\lambda\xi_\tau+\delta\,\xi_e\,\\[2mm]
  i\,\dot{\xi}_\tau=i\bm \sigma\cdot\bm \nabla\xi_\tau-\varepsilon\eta_e-\lambda\,\eta_\mu,\\[2mm]
  i\,\dot{\eta}_\tau=-i\bm \sigma\cdot\bm \nabla\eta_\tau+\varepsilon\xi_e+\lambda\,\xi_\mu,
  \end{array}\right.
\end{equation}
and then set $\delta=\varepsilon=\lambda$ later. By solving a typical algebraic equation
of third order, we find three eigenvalues of energy square being:
\begin{equation}\label{eq:26}
  \begin{array}{ll}
  E^2_i=p^2-m^2_i,\quad (i=1,2,3)\\[2mm]
  m^2_1=4\delta^2,\quad m^2_2=m^2_3=\delta^2\,.
  \end{array}
\end{equation}
The orthogonal transformation between mass and flavor eigenstates reads:
\begin{equation}\label{eq:27}
  \left(\begin{array}{ccc}\nu_1\\[3mm]\nu_2\\[3mm]\nu_3\end{array}\right)
  =\left(\begin{array}{ccc}\frac{1}{\sqrt{3}}&\frac{1}{\sqrt{3}}&\frac{1}{\sqrt{3}}\\[2mm]
  \frac{1}{\sqrt{6}}&\frac{1}{\sqrt{6}}&\frac{-2}{\sqrt{6}}\\[2mm]
  \frac{1}{\sqrt{2}}&\frac{-1}{\sqrt{2}}&0
  \end{array}\right)
  \left(\begin{array}{ccc}\nu_e\\[3mm]\nu_\mu\\[3mm]\nu_\tau\end{array}\right)\,.
\end{equation}
(We should write $m_1=2\delta$ and $m_2=m_3=-\,\delta$ in the proof.) Similar to the
procedure leading from (20) to (24), we get:
\begin{equation}\label{eq:28}
  \begin{array}{ll}
  |C_e(t)|^2= 1-\frac{8}{9}\sin^2[\frac{1}{2}(E_2-E_1)t],\\[2mm]
  |C_\mu(t)|^2=|C_\tau(t)|^2=\frac{4}{9}\sin^2[\frac{1}{2}(E_2-E_1)t]\,.
  \end{array}
\end{equation}
Some remarks are in order:

First, if neutrinos are propagating through a medium, the oscillation picture will be
modified drastically as discussed by Wolfenstein[19] and Mikheev-Smirnov[20]. The finite
length of wave packet must be considered[20, 3]. While different phase velocities ($u_p$)
lead to oscillations, different group velocities ($u_g$) result in a separation of the
wave packets in space. The situation becomes more complex because Eq.~(15) now means $u_g
>c$ in the superluminal case whereas $u_p<c$, implying that during the propagation, the
coherent superposition of different flavor states may be difficult to catch up the wave
packets which are separated faster than light. Hence I guess that the periodical flavor
transformation probability (28) must be strongly suppressed in a medium. Maybe the
oscillation becomes so incoherent that it actually provides a mechanism of detailed
balancing for neutrinos. So eventually an equally distributed population may be
established among three flavors. Could the experimental data of solar neutrino [2] be
explained in this way? I can only lay my hope on experts in this field.

Second, as stressed by Close[21], solving the solar neutrino problem qualifies as one of
the great moments in experimental science and neutrino physics is about to enter a golden
age. Now SNO experiment[2] and Kamland experiment in Japan actually have fixed three
flavors of neutrino and finally ruled out the existence of possible fourth flavor (or so
called 'sterile') neutrinos. Furthermore, by the K2K experiment in Japan[6] and some
other neutrino factories under construction or planning, physicists will be able to
detect neutrinos travelling a long distance on earth and measure their oscillations as
well as velocities. I was excited to learn the preliminary data of neutrino velocity
$v=3.0120481\times 10^8~{\rm m/s}$ along a distance 250 km by K2K in June 1999 [22]. So I
am anxious to await the new data as soon as possible.

Third, if we measure the mass square $m^2$ of neutrino with certain flavor when it is
just produced, like $\nu_e$ in (28), then $m^2$ must be an average of (mass)$^2$
eigenvalues with considerable standard deviation $\sigma$ due to the existence of
oscillation. For instance, in the above three-flavor model, we may predict a measured
value of $m^2$ being:
\begin{equation}\label{eq:29}
   m^2=<m^2>\pm\sigma=-\frac{8}{5}\,\delta^2\pm\frac{6}{5}\,\delta^2\,.
\end{equation}
But if the neutrino is an outcome of oscillation, like $\nu_\mu$ or $\nu_\tau$ in (28),
then its mass square $\widetilde{m}^2$ measured in flight is expected to be:
\begin{equation}\label{eq:30}
  \widetilde{m}^2=<\widetilde{m}^2>\pm\widetilde{\sigma}=-\frac{5}{2}\,\delta^2\pm\frac{3}{2}\,\delta^2\,.
\end{equation}
These two predictions are independent of the flavor in this model (Compare Eq.~(16)).

Fourth, if we modify all coupling terms in (25) into plus signs, then instead of (26), we
obtain:
\begin{equation}\label{eq:31}
  \begin{array}{ll}
  E^2_i=p^2+m^2_i,\quad (i=1,2,3)\\[2mm]
  m^2_1=4\delta^2,\quad m^2_2=m^2_3=\delta^2\,.
  \end{array}
\end{equation}
However, similar to (28), Eq.~(31) implies oscillation among three flavor states of
subluminal Dirac particles. But in fact, there is no any oscillation among electron, muon
and tau lepton. Hence we understand why the muon-electron transformation process like
$\mu\rightarrow e+\gamma$ is forbidden. This is because the flavor change term is
stemming from weak interactions with parity violation, if it is combined with the Dirac
equation (of $\mu$ or $e$) having parity conserved property, it would make the whole
equation asymmetric under the space inversion. In other words, it is the inconsistency of
symmetry that forbids the process of muon-electron transformation.

In summary, there are three things being intimately correlated together: As Dirac
particles obey the parity conservation law, they are bound to be subluminal and cannot
oscillate among different flavor states. By contrast, since neutrinos violate the parity
symmetry to maximum, they must be superluminal particles and are capable of oscillating
among three flavor states. In the later case, there is only one motion equation like (25)
shared by all three flavors of neutrino and antineutrino.

\begin{acknowledgments}
I thank S. Q. Chen, Z. X. Huang, P. Leung and D. Lu for bringing relevant references to
my attention and helpful discussions.
\end{acknowledgments}
\appendix
\section{Mass eigenstates of neutrino in two-flavor model}

Assuming the neutrino momentum $\bm p$ along $z$ axis and denoting
$\xi=\left(\begin{array}{ll}0\\1\end{array}\right)
\left[\eta=\left(\begin{array}{ll}1\\0\end{array}\right)\right]$ the left-handed
(right-handed) polarized spin state, we can express two normalized solutions of Eq.~(17)
for neutrino (with energy $E_{1,2}>0$) being:
\begin{widetext}
\begin{equation}\label{eq:A.1}
    \psi_1(z,t)=\frac{1}{\sqrt{V}}\,\theta_1(p)e^{i\,(p\,z-E_1 t)},\quad
  \theta_1(p)=\frac{1}{\sqrt{1+\Lambda^2}}\left(\begin{array}{cc}u_1(p)\\[2mm]\Lambda u_1(p)
  \end{array}\right),\quad
  u_1(p)=\sqrt{\frac{p+E_1}{2E_1}}\left(\begin{array}{cc}\xi\\[2mm]\frac{m_1}{p+E_1}\xi\end{array}
  \right),
\end{equation}
\begin{equation}\label{eq:A.2}
  \psi_2(z,t)=\frac{1}{\sqrt{V}}\,\theta_2(p)e^{i\,(p\,z-E_2 t)},\quad
  \theta_2(p)=\frac{1}{\sqrt{1+\Lambda^2}}\left(\begin{array}{cc}\Lambda u_2(p)\\[2mm] -u_2(p)
  \end{array}\right),\quad
  u_2(p)=\sqrt{\frac{p+E_2}{2E_2}}\left(\begin{array}{cc}\xi\\[2mm]\frac{m_2}{p+E_2}\xi\end{array}
  \right)\,.
\end{equation}
This is Eq.~(20). The WFs for antineutrino with positive energy $E_c=-E, (E=-E_{1,2})$
are:
\begin{equation}\label{eq:A.3}
  \psi_{c1}(z,t)=\frac{1}{\sqrt{V}}\,\theta_3(p)e^{-i\,(p\,z-E_1 t)},\quad  \theta_3(p)=\frac{1}{\sqrt{1+\Lambda^2}}\left(\begin{array}{cc} u_3(p)\\[2mm] \Lambda u_3(p)
  \end{array}\right),\quad
  u_3(p)=\sqrt{\frac{p-E_1}{2E_1}}\left(\begin{array}{cc}\eta\\[2mm]\frac{m_1}{p-E_1}\eta\end{array}
  \right)\,,
\end{equation}
\begin{equation}\label{eq:A.4}
  \psi_{c2}(z,t)=\frac{1}{\sqrt{V}}\,\theta_4(p)e^{-i\,(p\,z-E_2 t)},\quad
  \theta_4(p)=\frac{1}{\sqrt{1+\Lambda^2}}\left(\begin{array}{cc}\Lambda u_4(p)\\[2mm] -u_4(p)
  \end{array}\right),\quad
  u_4(p)=\sqrt{\frac{p-E_2}{2E_2}}\left(\begin{array}{cc}\eta\\[2mm]\frac{m_2}{p-E_2}\eta\end{array}
  \right)\,.
\end{equation}
\end{widetext}


\begin{thebibliography}{22}
\bibitem{1}Y. Fukuda, {\sl et al}. Evidence for oscillation of atmospheric neutrinos, Phys. Rev.
Lett. 81, 1562-1567(1998).
\bibitem{2}Q. R. Ahmad, {\sl et al}. Direct evidence for neutrino flavor transformation
from neutral-current interactions in the Sudbury Neutrino Observatory. Phys. Rev. Lett.
89, 011301-1-6(2002).
\bibitem{3}C. W. Kim and A. Pevsner, Neutrino in Physics and Astrophysics, (Harwood Academic Publoshers, 1993).
\bibitem{4}R. N. Mohapatra and P. B. Pal, Massive neutrinos in physics and astrophysics (World Scientific, 1991).
\bibitem{5}M. Goldhaber, L. Grozins and A. Sunyar, Helicity of neutrinos, Phys. Rev. 109, 1015-1017(1958).
\bibitem{6}H. Murayama, The origin of neutrino mass, Phys. World, May 2002, 35-39.
\bibitem{7}G. Farmelo, Edited, It must be beautiful--Great equations of modern science, (Granta Books,
London, NewYork, 2002).
\bibitem{8}D. J. Gross, Guage theory--past, present and future, talk delivered
at International Symposium in Honor of C. N. Yang's 70th Birthday, National Tsing Hua
University, Hsinchu, Taiwan, July 1992.
\bibitem{9}A. Chodos, A. I. Hauser and V. A. Kostelecky, The neutrino as a tachyon, Phys. Lett. B150,
431-435(1985).
\bibitem{10}E. Giannetto, G. D. Maccarrone, R. Mignani and E. Recami, Are muon neutrino
faster-than light particles? {\sl ibid}, 178, 115-119(1986).
\bibitem{11}J. Ciborowski and J. Rembielinski, Tritium decay and the hypothesis of tachyonic neutrinos, Europ.
Phys. J. C8, 157-161(1999).
\bibitem{12}T. Chang and G. J. Ni, An explanation of possible negative
mass-square of neutrini, Fizika B (Zagreb)11, 49-55(2002), hep-ph/0009291.
\bibitem{13}G. J. Ni and T. Chang, Two parameters describing a superluminal neutrino, Jour. of Shaanxi Normal
University (Nat. Sci. Edi.) 30(3) 32-39(2002), hep-ph/0103051.
\bibitem{14}G. J. Ni, There might be superluminal particles in
nature, {\sl ibid}, 29(3) 1-6(2001), hep-th/0201077; Superluminal paradox and neutrino,
{\sl ibid}, 30(4) 1-6(2002), hep-ph/0203060.
\bibitem{15}G. J. Ni, Evidence for neutrino being
likely a superluminal particle, submitted to Int. J. Mod. Phys. A, hep-ph/0206296.
\bibitem{16}G. J. Ni and S. Q. Chen, Advanced Quantum Mechanics,
(Chinese Edi: Press of Fudan University,2000); (English Edi: Rinton Press, 2002).
\bibitem{17}Particle Data Group, Review of Particle Physics, Euro. Phys. J. C15, 350(2000).
\bibitem{18}E. Sassaroli, Neutrino oscillation: a relativistic example of a two-level
system, Am. J. Phys. 67, 869-875(1999).
\bibitem{19}L. Wolfenstein, Neutrino oscillations in matter, Phys. Rev. D17, 2369-2374(1978).
\bibitem{20}S. P. Mikheev and A. Y. Smirnov, Resonance oscillations of neutrinos in matter, Soviet Phys.-USPEKHI 30, 759-790(1987).
\bibitem{21}F. Close, To catch a rising star, Reluctant heroes, New Scientist, 7 Dec. 33-36(2002).
\bibitem{22}Z. X. Huang, Modern advances in neutrino researches, Engineering Science (in
Chinese) 4(10), 90-93(2002).
\end{thebibliography}
\end{document}